\documentclass[twocolumn,english,showpacs,superscriptaddress,nofootinbib]{revtex4-1}
\usepackage[T1]{fontenc}
\usepackage[latin9]{inputenc}
\usepackage{geometry}
\usepackage{xcolor}
\usepackage{babel}
\usepackage{units}
\usepackage{braket}
\usepackage{amsmath}
\usepackage{amssymb}
\usepackage{stmaryrd}
\usepackage{graphicx}
\usepackage{wasysym}
\usepackage{bbold}
\usepackage[colorlinks=true,citecolor=magenta,linkcolor =blue]{hyperref}
\usepackage[all]{hypcap} 
\usepackage{times}

\makeatletter

\IfFileExists{lmodern.sty}{\usepackage{lmodern}}{}
\usepackage{babel}

\usepackage{hyperref}

\newcommand{\Duke}{Duke Quantum Center, Department of Physics and Electrical and Computer Engineering, Duke University, Durham, NC 27701}
\newcommand{\JQI}{Joint Quantum Institute and Joint Center for Quantum Information and Computer Science, University of Maryland and NIST, College Park, Maryland 20742}
\newcommand{\CSM}{Department of Physics, Colorado School of Mines, Golden, Colorado 80401}
\newcommand{\IonQ}{IonQ, Inc., College Park, MD 20740}
\newcommand{\MSU}{Department of Physics and Astronomy, Michigan State University, East Lansing, MI 48823}

\makeatother

\begin{document}
\title{Continuous Symmetry Breaking in a Trapped-Ion Spin Chain}
\date{\today}

\author{Lei Feng}
\email{lei.feng@duke.edu ; or.katz@duke.edu}
\affiliation{\Duke}
\thanks{These authors contributed equally.}

\author{Or Katz}
\thanks{These authors contributed equally.}
\affiliation{\Duke}

\author{Casey Haack}
\affiliation{\CSM}

\author{Mohammad Maghrebi}
\affiliation{\MSU}

\author{Alexey V. Gorshkov}
\affiliation{\JQI}

\author{Zhexuan Gong}
\affiliation{\CSM}

\author{Marko Cetina}
\affiliation{\Duke}

\author{Christopher Monroe}
\affiliation{\Duke}
\affiliation{\IonQ}

\begin{abstract}
One-dimensional systems exhibiting a continuous symmetry can host quantum phases of matter with true long-range order only in the presence of sufficiently long-range interactions. In most physical systems, however, the interactions are short-ranged, hindering the emergence of such phases in one dimension. Here we use a one-dimensional trapped-ion quantum simulator to prepare states with long-range spin order that extends over the system size of up to $23$ spins and is characteristic of the continuous symmetry-breaking phase of matter. Our preparation relies on simultaneous control over an array of tightly focused individual-addressing laser beams, generating long-range spin-spin interactions. We also observe a disordered phase with frustrated correlations. We further study the phases at different ranges of interaction and the out-of-equilibrium response to symmetry-breaking perturbations. This work opens an avenue to study new quantum phases and out-of-equilibrium dynamics in low-dimensional systems.
\end{abstract}
\maketitle

\begin{figure*}[t]
\begin{centering}
\includegraphics[width=16cm]{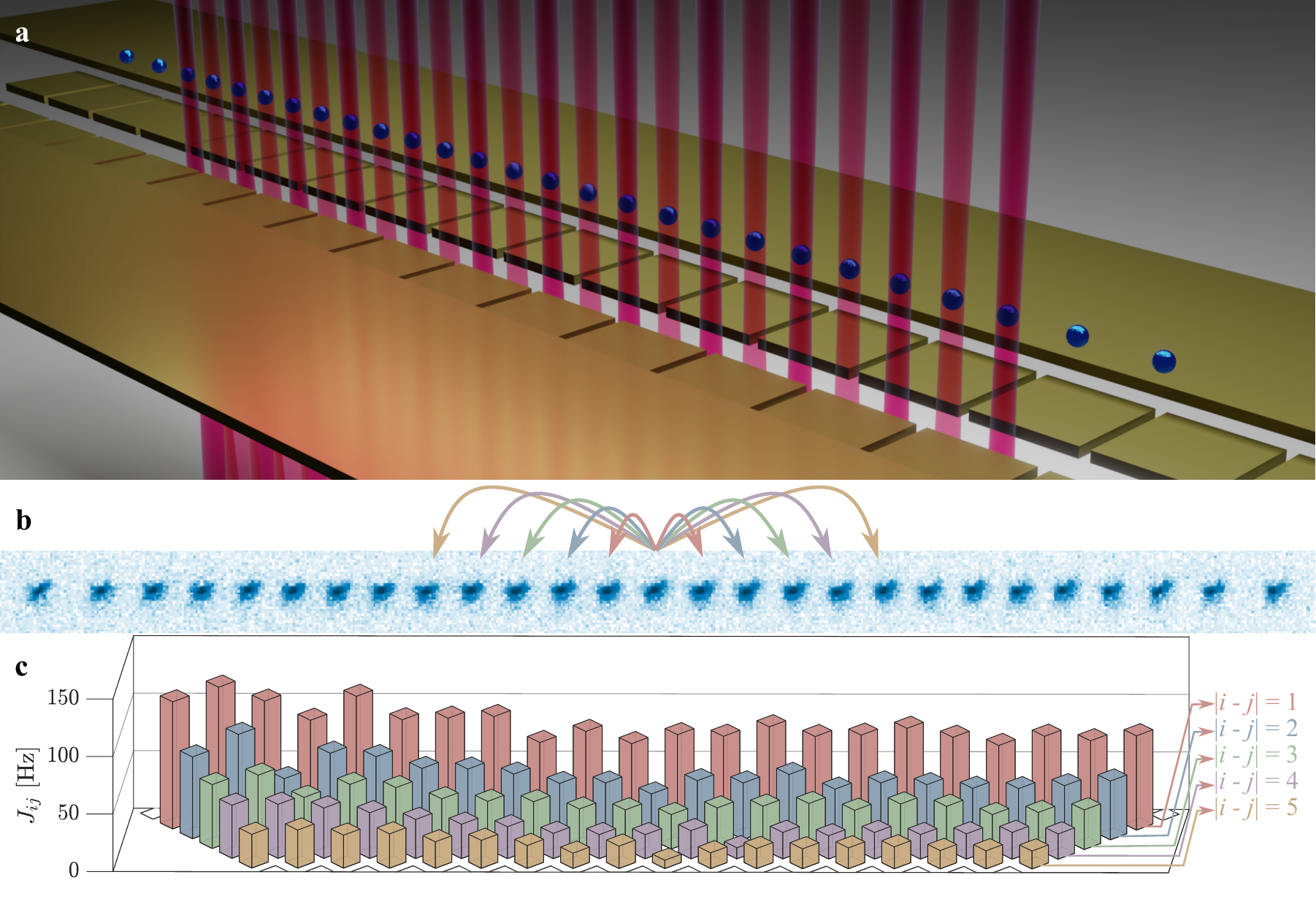}
\par\end{centering}
\centering{}\caption{\label{fig:Jij}\textbf{Trapped-ion crystal}. \textbf{a}, Illustration of a one-dimensional crystal of $27$ ions, confined in a linear Paul trap on a chip. A linear array of $23$ tightly focused and individually controlled laser beams simultaneously generates site-dependent fields and a programmable interaction between the trapped-ion spins; an additional beam,  propagating parallel to the trap surface, illuminates the entire ion chain from the side to facilitate these processes based on Raman-transitions (not shown). \textbf{b}, Fluorescence image of a crystal composed of 27 $^{171}$Yb$^+$ ions. \textbf{c}, Experimental reconstruction of the spin-spin interaction matrix $J_{ij}$ of the $23$ spins between the five nearest neighbors. The bars are horizontally aligned with the ion crystal image in \textbf{b}, and the colors indicate the interaction between spins at different distances $|i-j|$ (see Methods for the full modeled interaction).
}
\end{figure*}

\begin{figure*}[t]
\begin{centering}
\includegraphics[width=16cm]{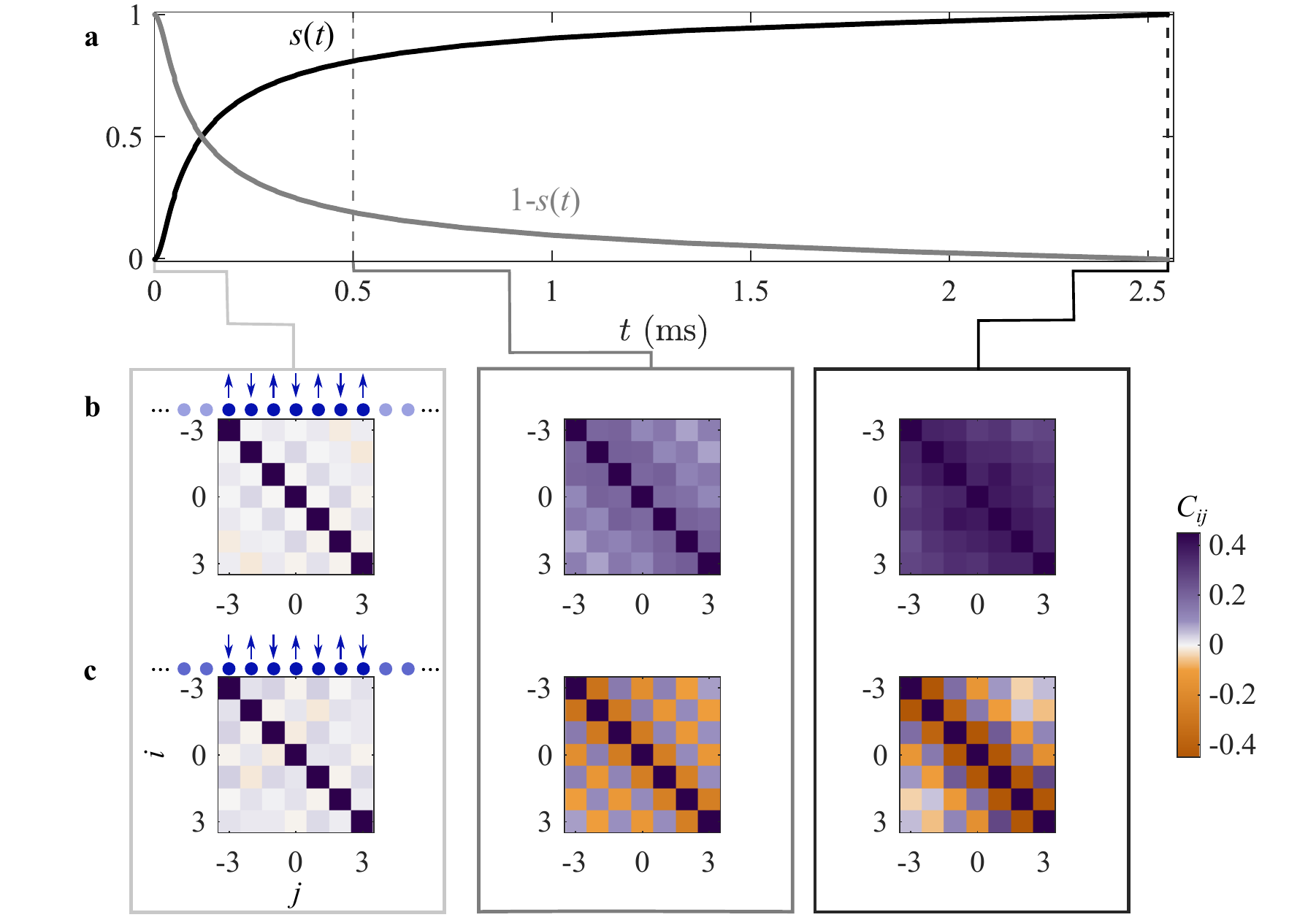}
\par\end{centering}
\centering{}\caption{ \textbf{Preparation of quantum phases.}  \textbf{a}, Adiabatic ramp profiles of the effective XY Hamiltonian, $s(t)$ and of the staggered magnetic field Hamiltonian, $1-s(t)$, as a function of time $t$.  \textbf{b,c}, Measured spin-spin correlations $C_{ij}=\langle \hat{\sigma}^{+}_i\hat{\sigma}^{-}_j+\hat{\sigma}^{-}_i\hat{\sigma}^{+}_j\rangle$ developed during the ramp for the subset of $N=7$ interacting spins ($-3\leq i,j \leq 3$) are indicated with small dark blue spheres; the other ions (light blue) are not addressed by optical fields and their spin states do not participate in the dynamics. \textbf{b}, Initializing the spins in the highest excited state of the staggered-field Hamiltonian along the $z$ direction leads to a low-temperature state of the ferromagnetic XY Hamiltonian at the end of the ramp. The positive correlations between all interacting spins indicate the continuous symmetry breaking (CSB) phase. \textbf{c}, Initializing the spins in the ground state of the staggered-field Hamiltonian prepares a low-temperature state of the antiferromagnetic XY Hamiltonian. 
\label{fig:temporal_evolution}}
\end{figure*}

\begin{figure*}[t]
\begin{centering}
\includegraphics[width=16.7cm]{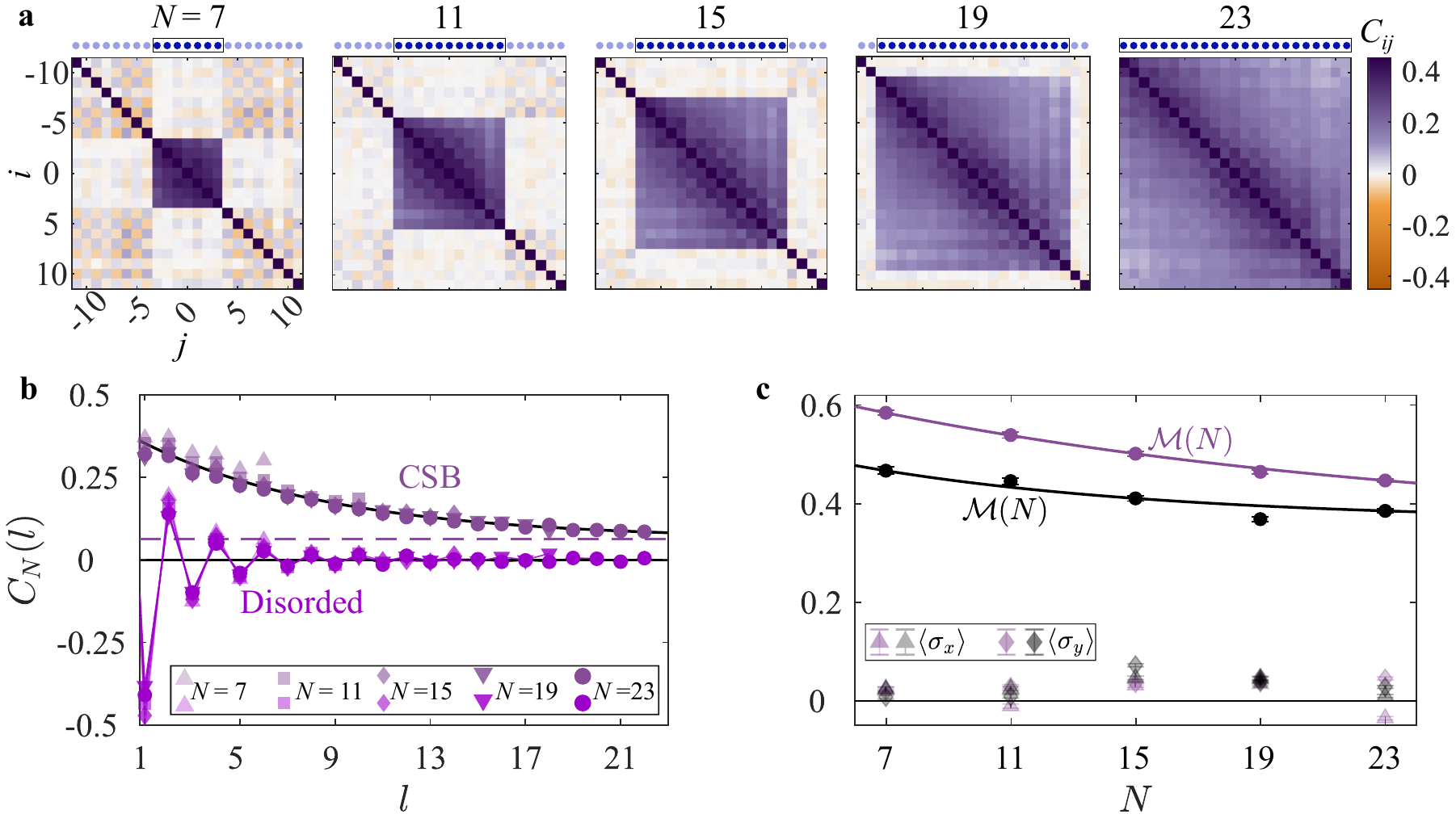}
\par\end{centering}
\centering{}\caption{ 
 \textbf{Long-range order}. \textbf{a}, The measured correlation matrix $C_{ij}$ of the prepared CSB phase for a subset of $N = 7,$ 11, 15, 19, and 23 interacting spins. Dark blue spheres indicate the ions that are illuminated by the  addressing beams. \textbf{b}, Spatially averaged correlations $C_N(l)=\frac{1}{N-l}\sum_j{C_{j,j+l}}$ as a function of the interaction distance $l$ for different subsystem sizes $N$. The correlations in the CSB phase (grayish purple) saturate asymptotically at a nonzero value of 0.062$\pm$0.005 in the $N,l\gg1$ limit (dashed line), manifesting long-range order. In contrast, the staggered correlations of the disordered phase (magenta) decay quickly to zero. The shape and brightness of the symbols indicate the number of interacting spins $N$. \textbf{c}, The purple data for the order parameter of the CSB phase $\mathcal{M}(N)$ [see Eq.~(\ref{eq:CSB_order})] saturates asymptotically at a sizeable nonzero value of 0.35$\pm$0.08, while the average transverse magnetization in the $x-y$ plane (pale symbols) is small, as expected from the continuous $U(1)$ symmetry in a finite system. Data in \textbf{a} and \textbf{b} as well as the purple data in \textbf{c} correspond to the interaction matrix that is partially shown in Fig.~\ref{fig:Jij}\textbf{c}. The black data in \textbf{c} corresponds to the interaction matrix that is partially shown in \ref{fig:Jij_alpha1d1} and that exhibits a shorter interaction range. The black data for the order parameter saturates asymptotically at 0.36 $\pm$ 0.18. Solid lines in \textbf{b, c} are fits to an exponential decay function with an additional offset. \label{fig:spatial_correlation}}
\end{figure*}

\begin{figure}[htbp]
\begin{centering}
\includegraphics[width=7.8cm]{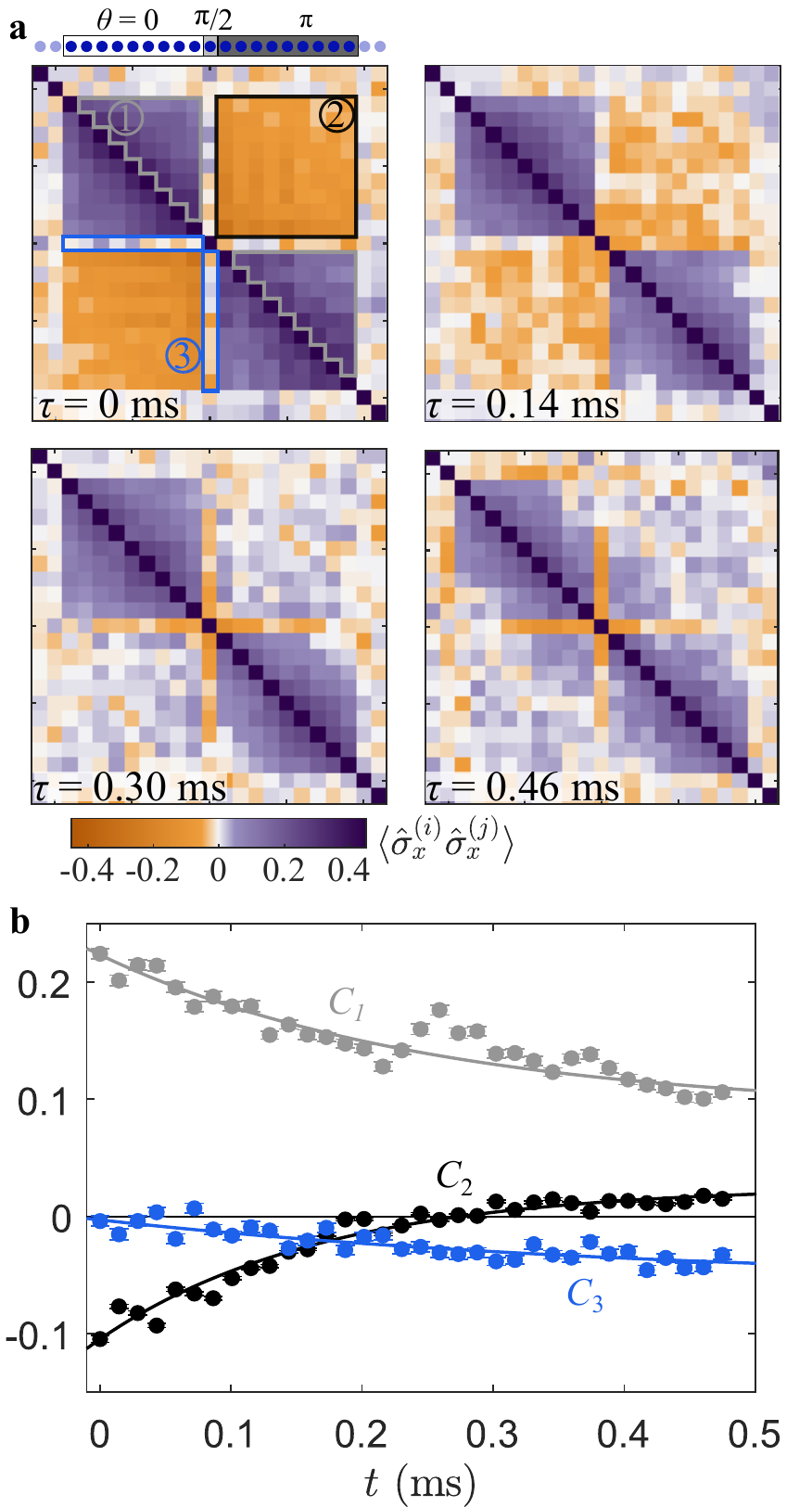}
\par\end{centering}
\centering{}\caption{\textbf{Out-of-equilibrium dynamics}.
Following the preparation of the CSB phase for $N$ = 19 spins, we perturb the state by rotating the spins in the $x-y$ plane by a spin-dependent angle $\theta$ at $\tau=0$. \textbf{a}, Measured spin-spin correlations $\langle\hat{\sigma}_x^{(i)}\hat{\sigma}_x^{(j)}\rangle$ developed during the evolution by the effective XY Hamiltonian for time $\tau$. At $\tau=0$ (top left), spins within the right or within the left side of the chain feature positive correlations, while correlations between spins on different sides are negative. The middle spin has a near-zero correlation with the rest of the chain. At $\tau=0.14$ ms (top right), the inter-correlations between the two sides decay quickly while the intra-correlations within each side are maintained. At later times (bottom) the entire chain develops positive correlations except for the middle spin which becomes anti-correlated to the rest of the chain. \textbf{b}, Average correlation as a function of time. The dots in different colors correspond to the correlation averaged within the corresponding colored contours shown in \textbf{a} (top left). Exponential fits with an offset (solid lines) are applied to guide the eye. The data in this figure corresponds to the interaction matrix partially shown in \ref{fig:Jij_alpha1d1}.\label{fig:quench}}
\end{figure}

The exploration of new phases of matter has long been a frontier of physics. Quantum phases are particularly interesting, featuring nonlocal and macroscopic properties that have no classical counterpart \cite{sachdev1999quantum}. One-dimensional quantum systems have captured special attention because they can often be efficiently described using various computational or analytic approaches.
\cite{giamarchi2003quantum,gong2016kaleidoscope,cazalilla2011one,chen2011classification}. The microscopic form and range of the interaction between constituent particles directly determine the macroscopic properties and phases that such systems can exhibit. Perhaps the best example is the Mermin-Wagner theorem \cite{mermin1966absence}, which forbids low-dimensional short-range interacting systems with a continuous symmetry from exhibiting 
long-range order at any finite temperature. 

One dimensional systems with long-range interactions, in contrast, can manifest phases with 
long-range order \cite{gong2016kaleidoscope,maghrebi2017continuous,haldane1983nonlinear,dalla2006hidden,gong2016topological,ren2022long,li2021long,herbrych2020block,giachetti2022berezinskii,potirniche2017floquet,patrick2017topological,bermudez2017long}. A prime example is a 
chain
of spin $1/2$ particles featuring long-range ferromagnetic interactions that have a continuous rotational 
$U(1)$ symmetry. Absent magnetic fields, the chain can possess an exotic phase where the spins exist in a superposition of collective states in the symmetry plane with no preferred orientation, yet due to the spontaneous breaking of the continuous symmetry, they host sizeable magnetic correlations across the entire chain 
\cite{gong2016kaleidoscope,maghrebi2017continuous}. Such a continuous symmetry-breaking (CSB) phase of matter has never been observed in a one-dimensional system.

Chains of trapped atomic ions are a pristine one-dimensional spin system, featuring high isolation from the environment, high-fidelity measurement and preparation of individual spins, and fully-connected spin-spin interactions whose strength and range can be controlled by optical fields \cite{Monroe2021,blatt2012quantum,joshi2022observing,zhang2017observation,morong2021observation,dumitrescu2022dynamical}. There have been 
proposals for observing CSB in trapped ion systems \cite{gong2016kaleidoscope,maghrebi2017continuous}, requiring simultaneous control over each optical field addressing individual ions in a long and closely-spaced crystal, which to date has been beyond experimental reach. 

Here we report on continuous symmetry breaking in a one-dimensional trapped-ion quantum simulator. Using simultaneous individual control of  a linear array of $23$ optical beams addressing individual ions, we prepare the system in a CSB phase, manifesting long-range spin-spin correlations. Individual control over the spins enables the precise engineering and measurement of the interactions between spins as well as the study of non-equilibrium dynamics under symmetry-breaking perturbations. These results represent a frontier in the control of 
quantum phases and open new avenues in studying low-dimensional quantum systems.

The trapped-ion crystal under study is comprised of twenty-seven $^{171}\textrm{Yb}^+$ ions confined in a linear Paul trap on a chip \cite{maunz2016high,egan2021fault,katz2022demonstration}, as illustrated in Fig.~\ref{fig:Jij}\textbf{a}. A fluorescence image of the crystal is shown in Fig.~\ref{fig:Jij}\textbf{b}. Each ion stores an effective spin comprised of two ``clock'' levels in its electronic ground-state ($\ket{\uparrow_z}\equiv \ket{F=1,M=0}$ and $\ket{\downarrow_z}\equiv\ket{F=0,M=0}$) \cite{Olmschenk2007}. We use a uniformly-spaced and array of tightly focused laser beams, together with an orthogonal wide global beam to simultaneously drive Raman transitions between the spin states of individual ions. The Raman addressing is sensitive to the motion along the wavevector difference between the individual and global addressing Raman beams \cite{Monroe2021}. 
The electrostatic trapping potential is configured to align the middle $23$ ions with the array of individual addressing beams. 
The two pairs of non-illuminated edge ions facilitate the alignment of the $23$ middle ions. The spins are initialized and measured using optical pumping and state-dependent fluorescence techniques \cite{Olmschenk2007} and the collective motional modes of the ion chain that mediate their interaction are cooled using sideband cooling \cite{egan2021scaling}. Single-spin rotations enable the orientation of each spin along any axis on the Bloch sphere for initialization or measurement.  

We deform the spin Hamiltonian as a function of time for different initial states to prepare different quantum phases of matter. Specifically, we ramp down a staggered transverse-field Hamiltonian and ramp up an effective long-range XY Hamiltonian \cite{Monroe2021} (see Methods), so that the total time-dependent Hamiltonian is
\begin{equation}\label{eq:total_Hamiltonian}
H=\frac{s}{2}\sum_{i<j}J_{ij}\left(\hat{\sigma}_+^{(i)}\hat{\sigma}_-^{(j)}+\hat{\sigma}_-^{(i)}\hat{\sigma}_+^{(j)}\right)+(1-s)\sum_{j}h_j\hat{\sigma}_z^{(j)},
\end{equation}
where $s=s(t)$ is a time-dependent parameter changing
from $0$ to $1$ during the time interval from $t=0$ to $t=T$ and $\boldsymbol{\hat{\sigma}}^{(j)}$ are the Pauli operators of the $j$th ion. Here $h_j=(-1)^{j}h$ is a uniform-magnitude magnetic field that alternates between adjacent spins. Each interaction amplitude $J_{ij}$ is positive and 
describes the flip-flop rate between the $i$th and $j$th spins. 

Simultaneous 
time-dependent control of the Raman beams enables the generation of the staggered-field Hamiltonian. This control also allows 
the selection of a subset of $N$ spins in the middle of the crystal that can interact with one another while remaining decoupled from the rest of the spins in the crystal: switching off the beam addressing the $n$-th ion nulls its hopping amplitude $J_{in}$ to all other ions $i$. The individual control also enables experimental reconstruction of the interaction matrix $J_{ij}$, as shown in Fig.~\ref{fig:Jij}\textbf{c} for the first five-nearest neighbors ($|i-j|\leq 5$). Here the measured long-range interaction decreases slowly 
as a function of the inter-spin spacing \cite{Monroe2021}, as modeled in the Methods section. The Hamiltonian evolution is also accompanied by decoherence induced by the optical drive, see Methods for details. 

To induce long-range correlations, we first initialize the spins in the N\'eel 
state in the $z$ basis, corresponding to the highest excited state of 
the staggered-field Hamiltonian. We then ramp the Hamiltonian with the profile of $s(t)$ shown in Fig.~\ref{fig:temporal_evolution}\textbf{a}. After the ramp, we immediately measure the transverse correlations ${C_{ij}=\langle\hat{\sigma}_+^{(i)}\hat{\sigma}_-^{(j)}+\hat{\sigma}_-^{(i)}\hat{\sigma}_+^{(j)}\rangle}$. We first consider the time evolution for a subset of $N=7$ interacting spins ($-3\leq i,j\leq 3$) shown in Fig.~\ref{fig:temporal_evolution}\textbf{b}. As the staggered field decreases and the interaction increases, correlations develop between all the interacting spins in the $x-y$ plane, indicating the CSB phase. On the other hand, when the spins are initialized in the ground state of the staggered-field Hamiltonian, shorter-range order develops in the $x-y$ plane after evolving under the same ramp \cite{gong2016kaleidoscope,maghrebi2017continuous}. Fig.~\ref{fig:temporal_evolution}\textbf{c} presents the formation of alternating and fast-decaying correlations between the $N=7$ interacting spins, indicating a disordered phase known as the XY phase \cite{gong2016kaleidoscope,maghrebi2017continuous}.

We focus on the CSB phase and study the correlations at the end of the ramp for a different number of interacting spins in the same ion chain, shown in Fig.~\ref{fig:spatial_correlation}\textbf{a}.
Dark-blue spheres indicate the set of interacting ions which are illuminated by the 
addressing beams. In all configurations, we observe sizeable and positive correlations $C_{ij}$ between the interacting spins. To quantify the spatial dependence of the long-range order, we present the spatially-averaged spin correlations $C_N(l)=\frac{1}{N-l}\sum_j{C_{j,j+l}}$ for different system sizes $N$ as a function of the interaction distance $1\leq l\leq N-1$ in Fig.~\ref{fig:spatial_correlation}\textbf{b} . The averaged correlations for different system sizes in the CSB phase nearly overlap and saturate to a nonzero value in the $N\gg1$ and $l\gg1$ limit indicated by the purple dashed line. In contrast, the spatially-averaged correlations of the disordered phase alternate in sign and quickly decay to zero.

We further quantify the averaged correlation of the CSB phase by extracting the order parameter \begin{equation}\label{eq:CSB_order}\mathcal{M}(N)=\sqrt{\frac{1}{N(N-1)}\sum_{i\neq j}C_{ij}},\end{equation}
as shown in Fig.~\ref{fig:spatial_correlation}\textbf{c}. The order parameter $\mathcal{M}(N)$ clearly saturates at a nonzero value, indicating the emergence of long-range order. On the other hand, the average spin magnetizations in the $x-y$ plane are nearly zero, as expected from the underlying $U(1)$ symmetry. The average magnetization along $z$ in the CSB phase, and the average magnetization in all three direction in the disordered phase are presented in \ref{fig:xy_csb_magnetization}.

The CSB phase is expected to persist 
in a system described by the XY Hamiltonian as long as the interactions have a sufficiently long range \cite{gong2016kaleidoscope,maghrebi2017continuous}. To verify this, we tune the optical Raman fields to repeat the experiment with shorter-range interactions corresponding to the experimentally-reconstructed interaction matrix $J_{ij}$ shown in \ref{fig:Jij_alpha1d1}. 
Following a similar protocol (see Methods), we prepare a spin state that exhibits long-range correlations, with a nonzero, yet smaller, order parameter for the CSB phase, shown in Fig.~\ref{fig:spatial_correlation}\textbf{c}. These results highlight the robustness of the CSB phase for different configurations and the key role played by the long-range interaction in realizing the emergent long-range order.  


The simultaneous individual control over the Raman fields provides a probe of the CSB phase's dynamical response to different perturbations. We observe the response of a perturbed CSB phase under the effective XY Hamiltonian in a system of $N = 19$ spins. We perturb the prepared CSB phase by rotating the spin of the individual ions by a variable angle $\theta_j$ about the $z$ axis while maintaining them in the $x-y$ plane. We invert the spins to the right of the center $(j>0)$  ($\theta_j=\pi$) while leaving the spins to the left of the center $(j<0)$ unperturbed ($\theta_j=0$). The central spin $(j=0)$ is rotated by $\theta_0=\pi/2$. 
This operation breaks the global $U(1)$ symmetry of the state while preserving the symmetry in the left and right subsystems.

Fig.~\ref{fig:quench} shows the measured correlations $\langle\hat{\sigma}_x^{(i)}\hat{\sigma}_x^{(j)}\rangle$ as a function of the evolution time $\tau$ from state preparation at $\tau=0$. Initially, the spins within each side ($i,j < 0$ or $i,j > 0$) of the crystal have positive correlations, while the correlations between spins on different sides are negative, as shown in Fig.~\ref{fig:quench}\textbf{a}. During the evolution, the inter-correlations between the two sides decay faster than the intra-correlations within each side. At longer evolution times, the two sides of the crystal overcome the perturbation and develop positive correlations, while the middle spin develops anti-correlations with the rest of the chain. In Fig.~\ref{fig:quench}\textbf{b}, we show the full time evolution of the system by plotting the averaged correlation $C_n=\sum_{\{i,j\}}C_{ij}$ for $\{i,j\}$  taken within the colored contours labeled as $n=1,2,3$ in Fig.~\ref{fig:quench}\textbf{a}. This demonstration shows our capability for further investigation 
of the properties of the symmetry-breaking phase.

In summary,
we observe a continuous symmetry-breaking phase with long-range order in a one-dimensional spin chain, manifested at different interaction ranges. Besides we show the preparation of a disordered phase with fast-decaying staggered correlations. As a teaser on the study of non-equilibrium dynamics, we show the full time evolution of the perturbed CSB phase. This work opens new avenues for studying quantum phases of matter in low-dimensional systems. 

\textit{Note added.}---While completing this project, we became aware of a complementary demonstration of CSB in a two-dimensional Rydberg array \cite{chen22b}.

\begin{acknowledgments}
This work is supported by the ARO through the IARPA
LogiQ program; the NSF STAQ, QLCI, RAISE-TAQS, and QIS programs; the DOE QSA
program; the AFOSR MURIs on Dissipation Engineering
in Open Quantum Systems and on Quantum Verification Protocols;  the ARO MURI on Modular Quantum Circuits; the DoE ASCR Quantum Testbed Pathfinder program; and the W. M. Keck Foundation.
\end{acknowledgments}

\renewcommand{\figurename}{}
\newcounter{extended_data_fig}
\setcounter{extended_data_fig}{1}
\renewcommand{\thefigure}{Extended Data Fig.~\arabic{extended_data_fig}}


\part*{\centerline{Methods}}

\section*{Interaction Hamiltonian}\label{sec:appendix_A}
We generate spin-spin interactions using Raman transitions that 
virtually excite collective motion of the ions. The beam that globally addresses the ion chain traverses an acousto-optical modulator (AOM) that is simultaneously driven with two radio-frequency (RF) signals, splitting the optical beam into two components with distinct tones. These two tones drive simultaneously the first red and blue sideband transitions in the dispersive regime with symmetric  detunings $\Delta$ of the Raman beatnote from the highest-frequency mode of the ion chain. We control the radial electrostatic potential to spectrally separate the two sets of radial modes and to align the wavevector difference of the Raman fields to the addressed set. In this configuration, we realize the Ising Hamiltonian $H_{\textrm{XX}}(t)=s(t)\sum_{ij}J_{ij}\hat{\sigma}_x^{(i)}\hat{\sigma}_x^{(j)}$. The time dependence of the Ising Hamiltonian is realized by varying the Rabi frequencies of the ions by a factor of $\sqrt{s(t)}$; this is achieved by controlling the power of $N\leq23$ RF signals feeding a multi-channel AOM which modulates the amplitude of the individually-addressing beams, while turning off all other $(23-N)$ channels.

We apply an effective transverse field at each spin by shifting the frequency of the beam addressing the $j$th ion as a function of time by $f_j=2 s(t)B+2(1-s(t))(-1)^{j}h$. This combination generates the transverse field Hamiltonian that is composed of two terms: a spatially-uniform transverse field ${H_{B}=s\sum_{j}B\hat{\sigma}_z^{(j)}}$ and a staggered field ${H_{h}=(1-s)\sum_{j}h(-1)^{j}\hat{\sigma}_z^{(j)}}$. 

The longer-range configuration with the interaction matrix in Fig.~\ref{fig:Jij}\textbf{c} corresponds to $\Delta\approx2\pi\times20$ kHz, $B=2\pi\times1.6$ kHz, $h=2\pi\times0.9$ kHz, and a ramp time of $T=2.55$ ms. The average nearest-neighbor interaction strength is $\bar{J}=\tfrac{1}{N-1}\sum_{i}J_{i,i+1}=2\pi\times0.09$ kHz. The second configuration with the interaction matrix in \ref{fig:Jij_alpha1d1} corresponds to $\Delta\approx2\pi\times55$ kHz, $B=2\pi\times6.5$ kHz, $h=2\pi\times4.2$ kHz, a ramp time of $T=0.54$ ms, and $\bar{J}=2\pi\times0.5$ kHz.

The applied transverse field overwhelms the Ising interaction because $B\gg \bar{J}$. Using the definition of the raising and lowering spin operators, $\hat{\sigma}_{\pm}^{(i)}=\tfrac{1}{2}(\hat{\sigma}_{x}^{(i)}\pm i\hat{\sigma}_{y}^{(i)})$ we can represent the Ising interaction in a frame rotating at the Larmor frequency of the uniform field by $\hat{\sigma}_{x}^{(i)}\hat{\sigma}_{x}^{(j)}\approx\tfrac{1}{2}(\hat{\sigma}_{+}^{(i)}\hat{\sigma}_{-}^{(j)}+\hat{\sigma}_{-}^{(i)}\hat{\sigma}_{+}^{(j)})$, bestowing fast oscillations to the $\hat{\sigma}_{\pm}^{(i)}\hat{\sigma}_{\pm}^{(j)}$ terms. This construction produces the effective XY Hamiltonian described in the main text. 

\vspace{10pt}
\section*{Experimental Reconstruction of the $J_{ij}$ matrix}\label{sec:appendix_B}
\begin{figure*}[t]
\begin{centering}
\includegraphics[width=16cm]{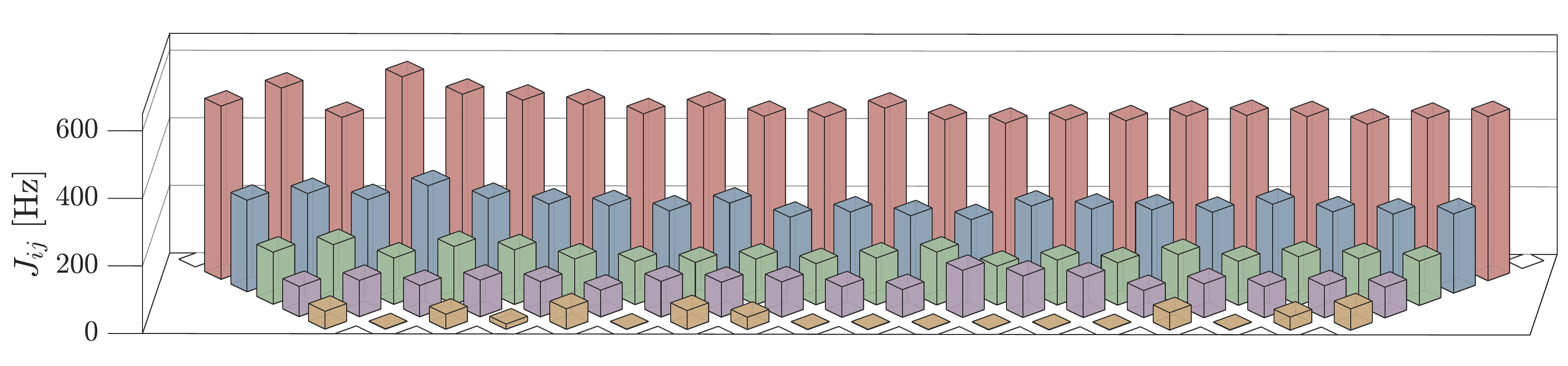}
\par\end{centering}
\centering{}\caption{\textbf{Partial reconstruction of the interaction matrix in the second experimental configuration}. The experimentally reconstructed $J_{ij}$ matrix is shown for a second experimental configuration up to five nearest neighbors. The full modeled interaction is detailed in the Methods section and is shown in Fig.~\ref{fig:chimera}. This matrix exhibits long-range interaction, which is nevertheless shorter than the interaction in Fig.~\ref{fig:Jij}\textbf{c}. \label{fig:Jij_alpha1d1}}
\end{figure*}

\begin{figure*}[t]
\begin{centering}
\includegraphics[width=17cm]{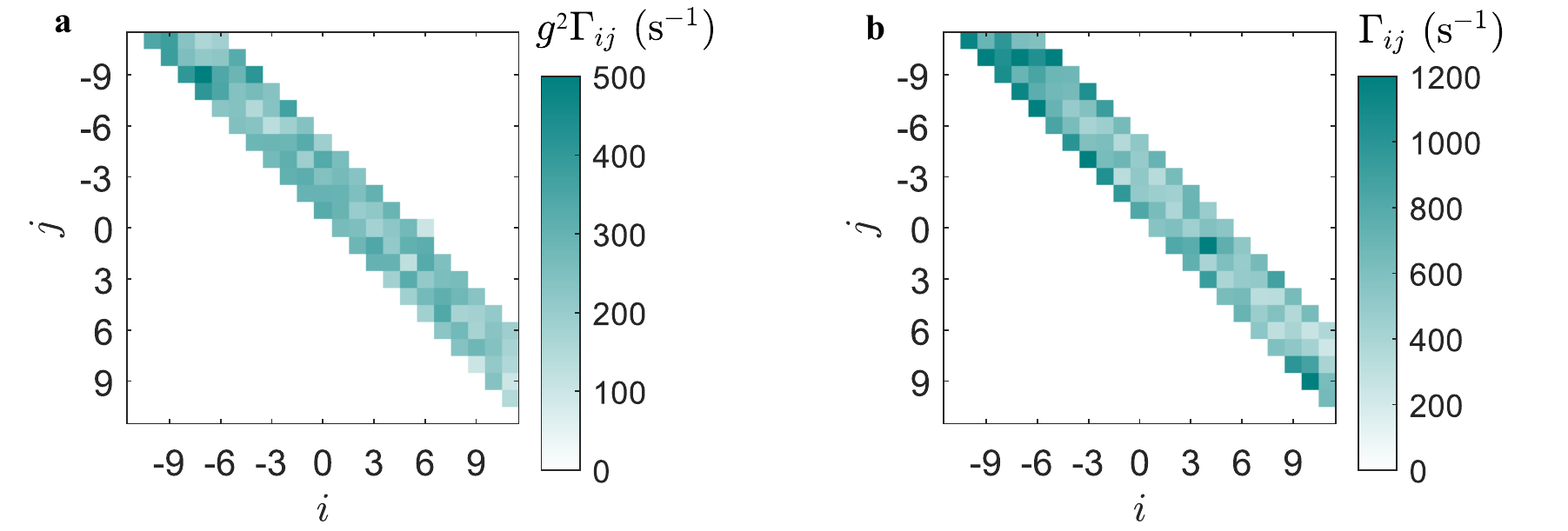}
\par\end{centering}
\centering{}\caption{ 
 \textbf{Decoherence rate matrix}. The measured decoherence rate matrix $\Gamma_{ij}$ is extracted from the reconstruction protocol for the up to five nearest neighbors (see Methods). \textbf{a}, The measured relaxation accompanying the interaction matrix in the first configuration (Fig.~\ref{fig:Jij}\textbf{c}) with $g=1.3$. \textbf{b}, The measured relaxation accompanying the interaction matrix in the second configuration (Fig.~\ref{fig:Jij_alpha1d1}). \label{fig:decoherence}}
\end{figure*}

\begin{figure*}[t]
\begin{centering}
\includegraphics[width=17cm]{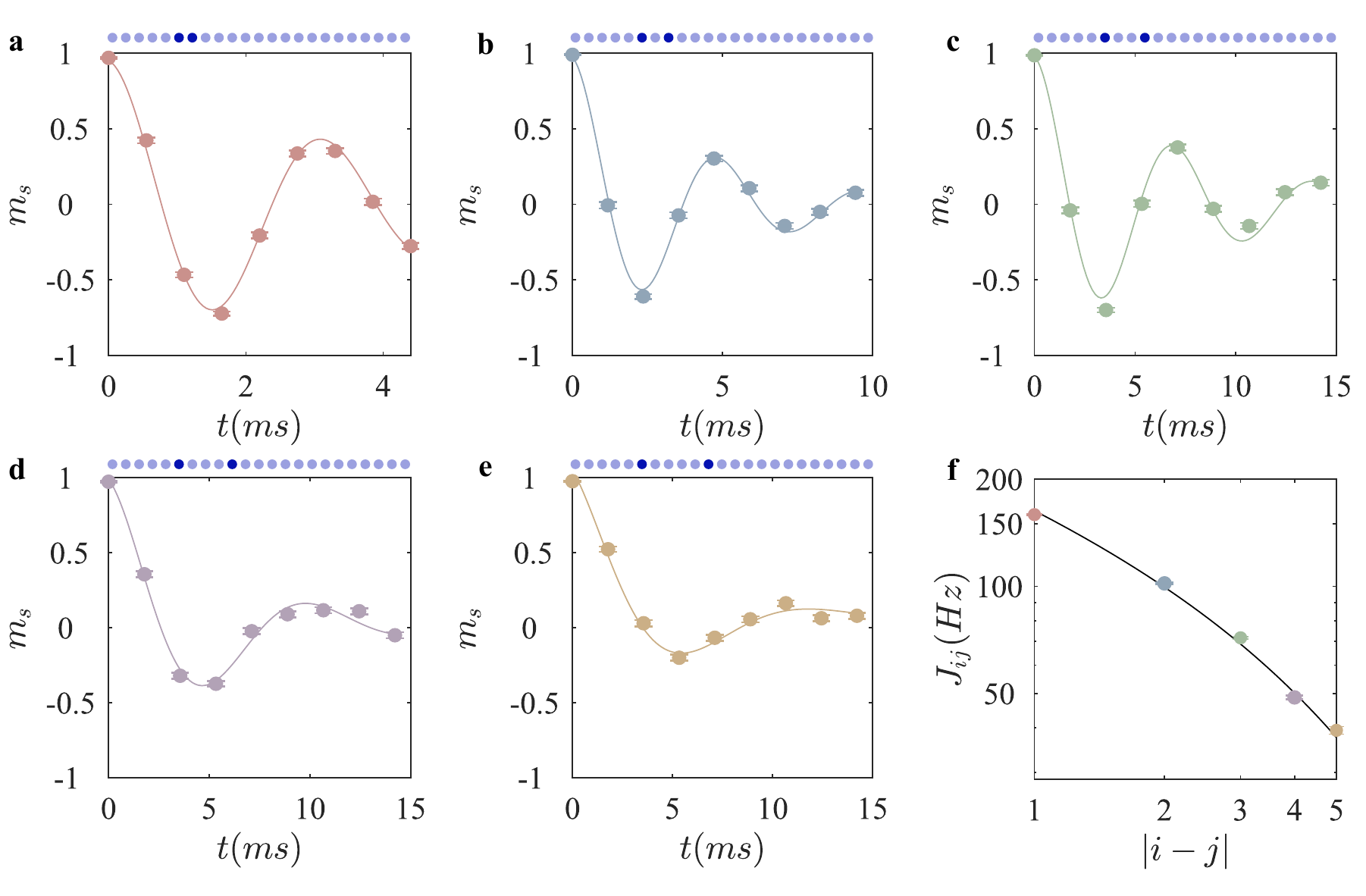}
\par\end{centering}
\centering{}\caption{ 
 \textbf{Demonstration of the reconstruction protocol}. We measure the interactions between the $i$~=~-6 ion with its up to five-nearest neighbors, by turning on the a single pair of beams addressing two ions a time. Specifically $(i,j)$ = (-6,-5) in \textbf{a}, (-6,-4) in \textbf{b}, (-6,-3) in \textbf{c}, and (-6,-2) in \textbf{d} and  (-6,-1) in \textbf{e}, as indicated by a dark blue sphere. We fit the staggered magnetization $m_s = \tfrac{1}{2}\langle\sigma_z^i-\sigma_z^j\rangle$ to the function $y = \cos(\pi g^2J_{ij}t)e^{-g^2\Gamma_{ij}t}$ to extract the interaction strength $J_{ij}$ and the decoherence rate $\Gamma_{ij}$. The interaction rate as a function of the inter-ion spacing is shown in \textbf{f}. The fitted $J_{ij}$ for $i=-6$ (circles). The black line corresponds to the fit function in Eq.~(\ref{eq:chimera}) with fitting parameters $\alpha'=0.44,\,\beta'=0.19$. The exemplary data in this figure corresponds to the interaction matrix in Fig.~\ref{fig:Jij}c with a scaling factor $g=1.3$; see Methods. \label{fig:Jij_example}}
\end{figure*}

 We measure each $J_{ij}$ element by turning on the two beams addressing the $i$th and $j$th ions while turning off all other beams in the array. The ions are initialized in the state $\ket{\uparrow_z^{(i)}\downarrow_z^{(j)}}$ for $j>i$, and adjust the transverse field to zero ($f_i=f_j=0$). We apply a constant-amplitude pulse with a Rabi frequency that is scaled by $g=1.3$ in the first configuration (interaction matrix in Fig.~\ref{fig:Jij}) and by $g=1$ in the second configuration (interaction matrix in \ref{fig:Jij_alpha1d1}) and measure the population oscillations. We fit the average staggered magnetization $\tfrac{1}{2}\langle\hat{\sigma}_z^{(i)}-\hat{\sigma}_z^{(j)}\rangle$  to the function $\exp{(-g^{2}\Gamma_{ij} t)}\cos(\pi g^{2} J_{ij}t) $ using $J_{ij}$ and $\Gamma_{ij}$ as fitting parameters. The measured values of $\Gamma_{ij}$ are given in \ref{fig:decoherence}, and an example of the reconstruction is shown in \ref{fig:Jij_example}.

\stepcounter{extended_data_fig}

\vspace{10pt}
\section*{Numerical calculation of the $J_{ij}$ matrix}\label{sec:appendix_C}
We calculate the interaction matrix $J_{ij}$ that results from applying a spin-dependent optical dipole force with the Raman lasers, following Refs.~\cite{kim2009entanglement,Monroe2021}. These lasers, generate coupling between the spins and the collective motional modes along a single radial direction, virtually exciting phonons that mediate
the spin-spin interaction 
\begin{equation} \label{eq:Jij_formula}J_{ij}=\sum_{k}\frac{\eta_{ik}\eta_{jk}\Omega_{i}\Omega_{j}}{2(\Delta+\omega_{1}-\omega_{k})}.
\end{equation}
The spin-motion coupling matrix is represented by the Lamb-Dicke parameters $\eta_{nk}=0.08b_{nk}$, where $b_{nk}$ is the normal model matrix element describing the coupling between spin $n$ and motional mode $k$ \cite{katz2022demonstration}. We numerically calculate the matrix $b_{nk}$ and the frequencies of the motional modes $\omega_k$, listed in decreasing order, for the applied trapping potentials; we consider a quadratic trapping potential in the radial direction with center-of-mass frequency $\omega_{1}=2\pi\times3.3$ MHz and an axial potential of $V(x)=250\times x^{4}-0.1\times x^{2}$ where $x$ is the coordinate along the chain axis in millimeters and $V$ is the axial electrostatic potential in electron volts. This potential yields a nearly uniform-spaced ion chain for the inner 23 ions with a spacing of $3.75\,\mu$m. $\Omega_{i}$ represents the equivalent resonant carrier Rabi frequency at ion $i$, and we assume a spatially uniform profile.  

\begin{figure*}[t]
\begin{centering}
\includegraphics[width=8.7cm]{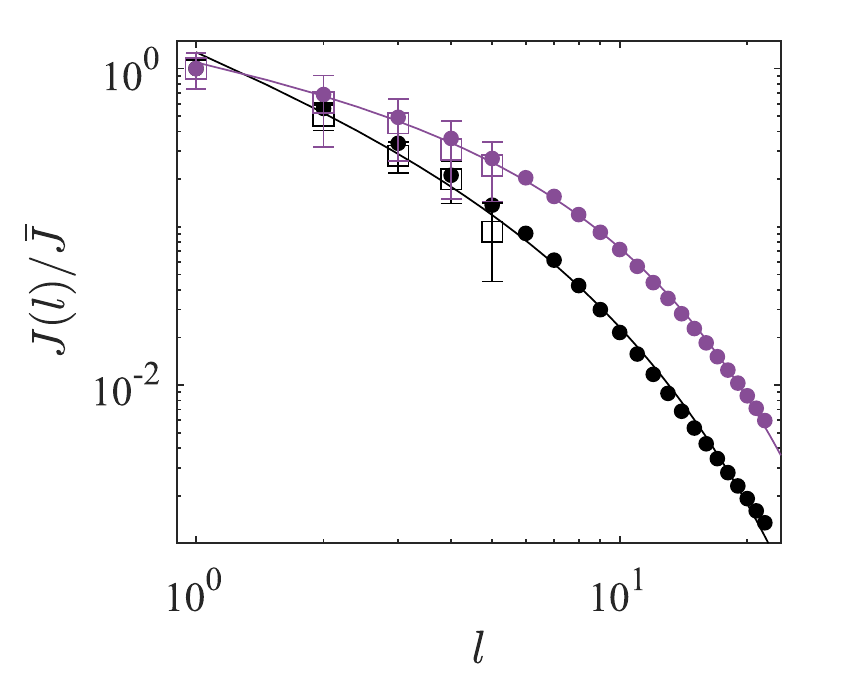}
\par\end{centering}
\centering{}\caption{ 
 \textbf{Modeled long-range interaction}. We calculate the spin-spin interaction based on a simple model of the trap potential for the two experimental configurations. The filled circles indicate the numerically calculated values with no free parameters. The solid lines are fits to the numerical results with a profile of $J(l) = \bar{J}e^{-\beta'(l-1)}l^{-\alpha'}$. The fitted parameters are $\alpha'=0.44,\,\beta'=0.19$ for the first configuration (purple) and $\alpha'=1,\,\beta'=0.19$ for the second configuration (black). Open squares are the experimental data in the two experimental configurations, where the bars represent the spread of measured values of all pairs at a specific spacing $l$.\label{fig:chimera}}
\end{figure*}

In \ref{fig:chimera} we present the numerically calculated averaged interaction $J(l)=\frac{1}{N-l}\sum_{i}J_{i,i+l}$ as a function of the distance $l$ for the two configurations (circles), where $\bar{J}\equiv J(1)$. We also present the average experimentally-measured interaction (open squares) where the bars reflect the total spread of values between different pairs, excluding points for which the error in the reconstructed value exceeded the actual measured value (only for several elements with $l=5$ in the second configuration which appear in \ref{fig:Jij_alpha1d1} as zero). The agreement between the measured value and the calculated values are in a good agreement. We fit the theoretical values to the fitting function
\begin{equation} \label{eq:chimera}J(l) = \bar{J}e^{-\beta'(l-1)}l^{-\alpha'},
\end{equation}
which we adapt from Ref.~\cite{pagano2020quantum}. The fitted parameters are $\alpha'=0.44,\,\beta'=0.19$ for the first configuration (purple line) and $\alpha'=1,\,\beta'=0.19$ for the second configuration (black line).

\stepcounter{extended_data_fig}

\stepcounter{extended_data_fig}

\begin{figure*}[t!]
\begin{centering}
\includegraphics[width=17cm]{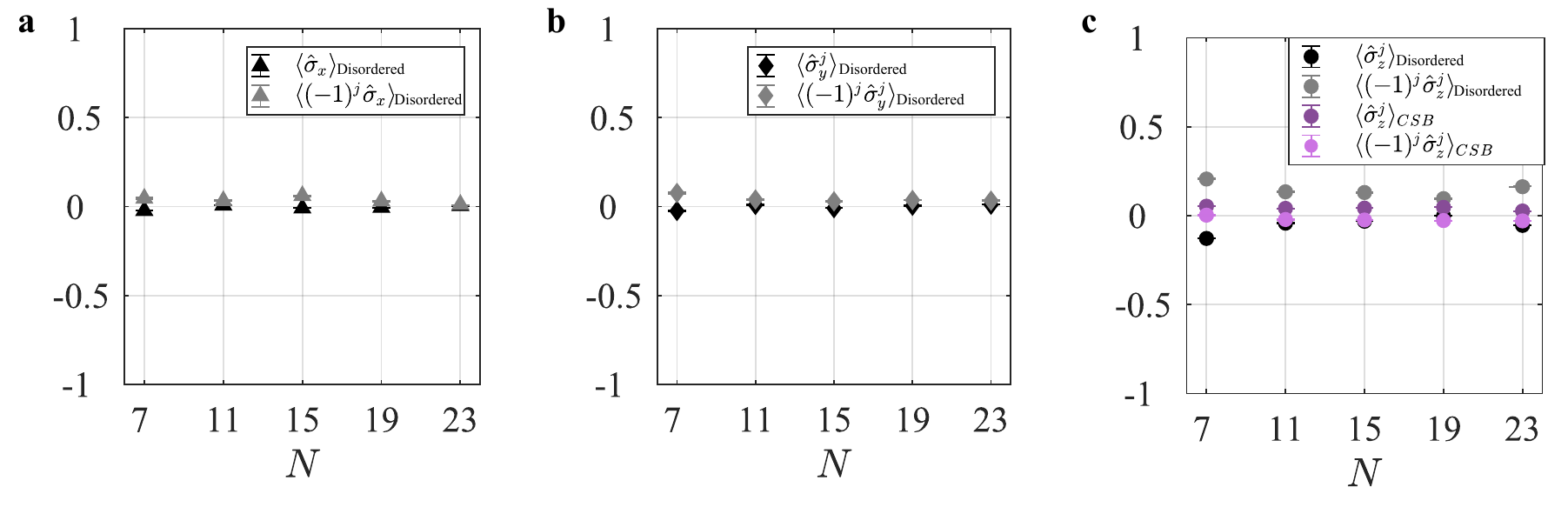}
\par\end{centering}
\centering{}\caption{ 
 \textbf{Average magnetization}. The measured regular and staggered magnetization along the $x$ (\textbf{a}), $y$ (\textbf{b}), and $z$ (\textbf{c}) axes for the first configuration (with interaction matrix in Fig.~\ref{fig:Jij}\textbf{c}).
 \label{fig:xy_csb_magnetization}}
\end{figure*}
\stepcounter{extended_data_fig}

\stepcounter{extended_data_fig}

\clearpage
\bibliography{Refs}

\end{document}